\documentclass[ aps,prl, amsmath,amssymb,reprint]{revtex4-1}

\pdfoutput=1

\usepackage{graphicx}% Include figure files
\usepackage{dcolumn}% Align table columns on decimal point
\usepackage{bm}% bold math

\begin{document}

\title{Quasistatic and Pulsed Current-Induced Switching with Spin-Orbit Torques in Ultrathin Films with
Perpendicular Magnetic Anisotropy} %Title of paper

\author{Yu-Ming Hung}
\email[]{yuming.hung@nyu.edu}
\author{Laura Rehm}
\author{Georg Wolf}
\author{Andrew D. Kent}

%\homepage[]{Your web page}
%\thanks{}
\affiliation{Department of Physics, New York University, New York, New York 10003, USA}

\date{\today}

\begin{abstract}
Spin-orbit interaction derived spin torques provide a means of reversing the magnetization of perpendicularly magnetized ultrathin films with currents that flow in the plane of the layers. A basic and critical question for applications is the speed and efficiency of switching with nanosecond current pulses. Here we investigate and contrast the quasistatic (slowly swept current) and pulsed current-induced switching characteristics of micron scale Hall crosses consisting of very thin ($<1$ nm) perpendicularly magnetized CoFeB layers on $\beta$-Ta. While complete magnetization reversal is found at a threshold current density in the quasistatic case, short duration ($\leq 10$ ns) larger amplitude pulses ($\simeq 10 \times$ the quasistatic threshold current) lead to only partial magnetization reversal and domain formation. We associate the partial reversal with the limited time for reversed domain expansion during the pulse.
\end{abstract}


%\pacs{}% insert suggested PACS numbers in braces on next line

\maketitle 
\section{Introduction}
In heavy metal-ferromagnet thin film heterostructures, current flowing in the plane of the layers can induce magnetization switching through spin-orbit interactions \cite{Miron,Liu1}. This is particularly interesting for ferromagnetic layers with perpendicular magnetization because their large magnetic anisotropy permits very stable magnetic states, even in elements that are nanometer scale in lateral dimension ($\sim 20$ nm diameter, see, for example \cite{Kent2015}). Further, spin-orbit torque switching enables three terminal memory elements with separate write and read current paths \cite{LiuSci,Pai,Cubu} as well as new types of spin-based logic devices~\cite{Datta}. A number of heavy metals have been shown to produce large spin-orbit torque to current ratios, including $\beta$-phase Ta and W as well as Pt, and thus lead to relatively low current densities for magnetization switching \cite{Hao2015}. However, with a few notable exceptions \cite{Garello,LoConte}, experimental studies have focused on quasistatic magnetization switching characteristics, i.e. switching for slowly varying (nearly dc) currents. It is clearly of interest for applications and basic understanding to explore the dynamics of magnetization switching for short current pulses.

Here we present a comparison of the quasistatic and the dynamic switching characteristics of high quality $\beta$-Ta/CoFeB/MgO heterostructures with large perpendicular magnetic anisotropy. We find full magnetization reversal at a threshold current for slowly swept current with applied in-plane magnetic fields. The scalar product of the current density and the applied field is shown to determine the sense of the magnetization switching, i.e. whether switching is from magnetized up to down or vice-versa \cite{Park}. However, only partial switching is observed for sub-$10$ ns current pulses of amplitude 10 times the quasistatic threshold current. We suggest that the origin of the partial switching is the limited time for domain expansion during the pulse.
\begin{figure}[!b]
\centering
\includegraphics[width=3.4in]{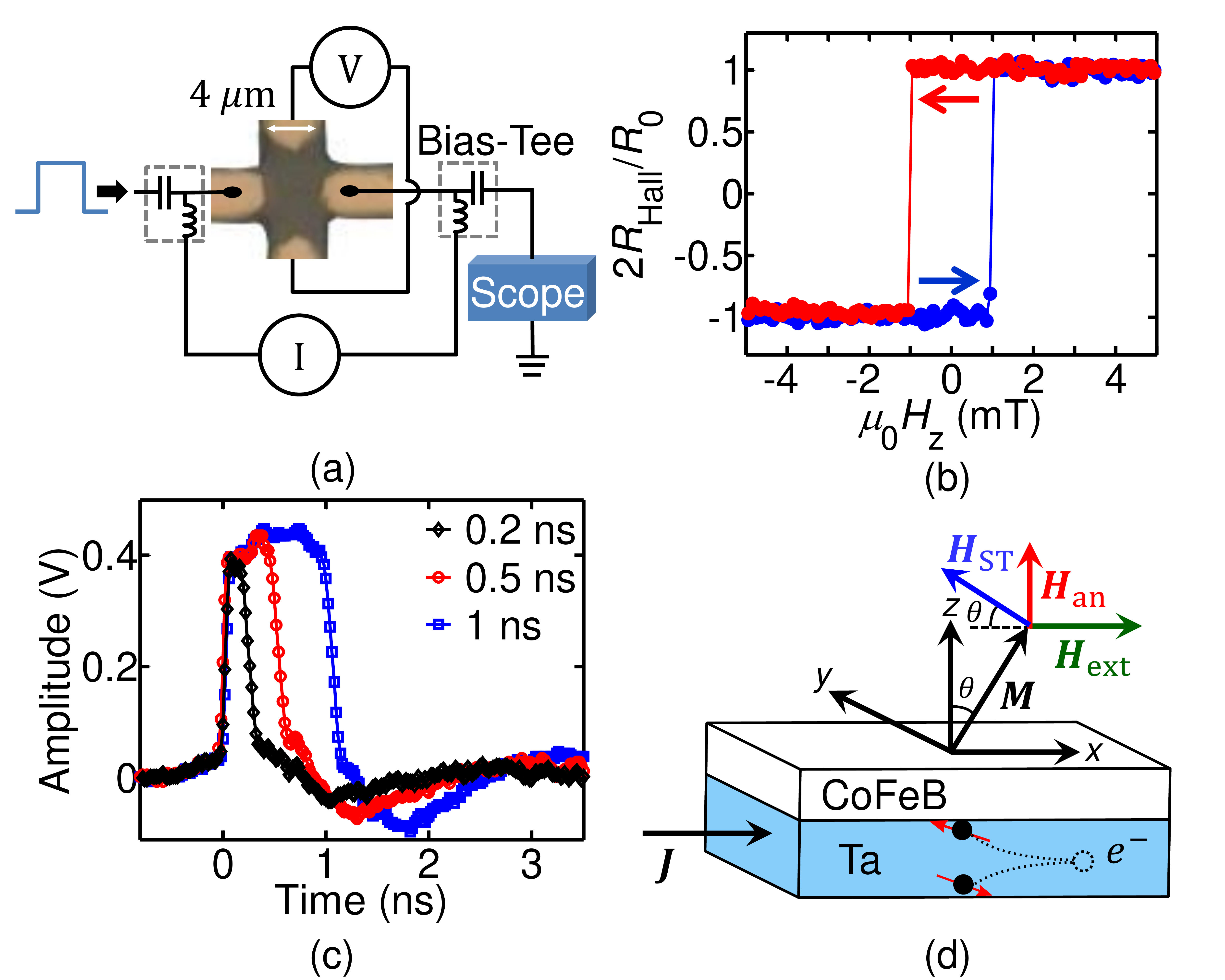}
\caption{(a) Measurement circuit for the quasistatic and pulse-current experiments with an optical image of the sample. (b) Hall resistance vs. out-of-plane field showing the up state ($2R_{\textrm{Hall}}/R_0 = 1$), down state ($2R_{\textrm{Hall}}/R_0 = -1$), and a coercive field of $H_{c}=1$ mT. $R_0 = 2.02\ \Omega$. (c) Pulse transmission measured with an oscilloscope. The pulse rise time and fall time (20 \% - 80\%) are 56 and 102 ps, respectively. (d) Schematic showing anisotropy ($\textit{\textbf{H}}_{\textrm{an}}$), external ($\textit{\textbf{H}}_{\textrm{ext}}$), and effective field ($\textit{\textbf{H}}_{\textrm{ST}}$) from spin-orbit torques.}
\label{fig_circuit}
\end{figure}
\section{Experiments Methods}
We start with layers grown by magnetron sputtering on oxidized Silicon wafers using a Singulus deposition system.%\cite{Singulus}. 
 The layer stack is Substrate$| 5$ $\beta -$Ta$| 0.8$ Co$_{0.4}$Fe$_{0.4}$B$_{0.2} |1.6$ MgO $|2$ Ta with the numbers indicating the layer thicknesses in nanometers. The $2$ nm Ta top layer serves to protect the sample from oxidization. The samples are annealed at 300~$^{\circ}$C for 2 hours to crystallize the CoFeB and the presence of the $\beta$-phase of Ta was verified by x-ray diffraction. Ferromagnetic resonance spectroscopy was used to determine the effective magnetization of the CoFeB layer defined as the perpendicularly anisotropy field $2K_u/M_s$ minus the demagnetization field, $\mu_0M_{\textrm{eff}}=2K_u/M_s-\mu_0M_{s}$. We found $\mu_0M_{\textrm{eff}}\approx 0.6$~T, indicating a strong perpendicular anisotropy, much larger than demagnetization field, and a magnetic easy axis perpendicular to the film plane. These results are consistent with earlier studies showing a perpendicular magnetic interface anisotropy associated with the CoFeB$|$MgO and CoFeB$|$Ta interfaces \cite{Ikeda,Worledge2012}.

Hall bar structures were then fabricated from these films using optical lithography and ion milling. An optical microscope image is shown in Fig.~\ref{fig_circuit}(a). The width of the arms of Hall crosses were between 2 to 8 $\mu$m and a SiO$_2$ dielectric protective layer was deposited in-situ directly after ion milling. Electric contacts consisting of Ta$|$Cu were deposited directly on the CoFeB in the arms of the Hall cross after removing the MgO barrier. We note that the CoFeB and $\beta$-Ta are in the same micron scale Hall bar shape. The results presented here were acquired on a 4 $\mu$m cross device. More than 15 devices have been measured and show similar characteristics. All the experiments were done at room temperature.

Figure~\ref{fig_circuit}(a) shows the measurement circuit for quasistatic and pulse-current experiments. We determine the magnetization state using the anomalous Hall effect by measuring the Hall resistance $R_{\textrm{Hall}}$. The anomalous Hall effect is a measure of the average z-component of the magnetization in the area in which current flows. Figure~\ref{fig_circuit}(b) shows the Hall resistance as a function of field applied perpendicular to the film plane $H_z$. A square hysteresis loop is seen with sharp jumps in the Hall resistance at \mbox{$\pm 1$ mT} where the magnetization reverses, which is the coercive field of the CoFeB, i.e. \mbox{$\mu_0 H_c=1$ mT}.  This small coercive field ($H_c \ll M_\textrm{eff}$) is indicative of magnetization reversal that proceeds by reversed domain nucleation (for $H_z \simeq H_c$) and growth \cite{Dafine}.

To determine the switching behavior of the sample in the quasistatic case the current was routed via the low frequency ports of the bias tees, as shown in Fig.~\ref{fig_circuit}(a). Measurements were conducted by sweeping the current at a fixed in-plane oriented field in the $x$ direction. Pulsed current was injected using a Picosecond lab pulse generator and the pulse current transmission was measured with a 20 GHz bandwidth Tektronix real time oscilloscope, which provided a 50 $\Omega$ termination to a coplanar waveguide structure that contains the Hall cross. The transmission of short pules is shown in Fig.~\ref{fig_circuit}(c), indicating a rise time of 56 ps  (20 to 80\%) and small negative overshoot after the pulse. To determine the pulse switching probability (defined below), pulses with duration between 200~ps and 10~ns and amplitudes up to 83.5 MA/cm$^2$ were injected in the presence of an external in-plane field. An initial fully saturated state is prepared using large dc current and in-plane field. Complete saturation of the film was verified with magneto-optic Kerr imaging (MOKE) using the same measurement protocol (images not shown). With a small dc current ($J$ = 0.125 MA/cm$^2$), we measure the Hall resistance to determine the magnetization state both before and after the pulse injection and thus determine whether or not magnetization reversal has been triggered by the pulse.
\section{Results and Discussion}
\begin{figure}[!t]
\centering
\includegraphics[width=3.4in]{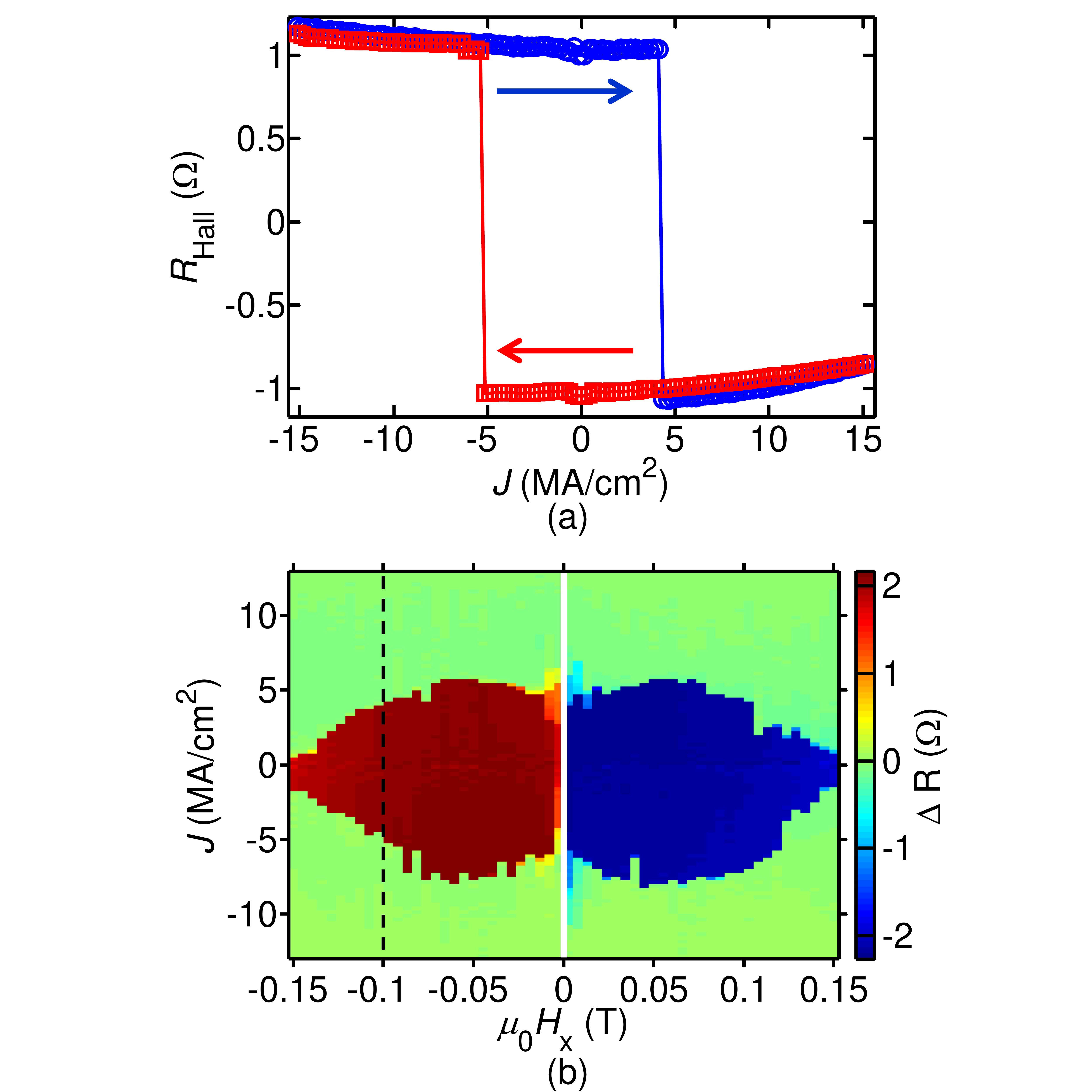}
\caption{(a) Quasistatic current-induced magnetization switching with in-plane applied field $\mu_0 H_{x}$ =  -100 mT and a current sweep rate of 50 $\mu$A/ms. (b) State diagram of a 4 $\mu$m Hall cross. The resistance obtained for increasing current is subtracted from the resistance obtained for decreasing current to display the bistable region. The black dash line is the cut showing in (a).}
\label{fig_DC}
\end{figure}
Figure~\ref{fig_DC}(a) shows a measurement of the Hall resistance for a slowly swept current with a fixed in-plane applied field \mbox{($\mu_0 H_{x}$ = -100 mT)}. The current density to switch from magnetized up to down is 4.25 MA/cm$^2$ and from down to up is -5.3 MA/cm$^2$. Repeating this measurements as a function of in-plane field ($10\leq\left|\mu_0 H_x\right|\leq150$ mT), we obtain the state diagram in Fig. \ref{fig_DC}(b), with the color representing the resistance for increasing current minus the resistance for decreasing current.

Thus the red and blue colors illustrate the bistable region: the parameter range for which both up and down magnetic states are possible. Different colors to the left and right of $H_x=0$ show clearly that the field polarity determines the sense of the magnetization switching, i.e. with current ramping up slowly from negative to positive, negative applied field leads to switching from magnetized up to down while positive applied field leads to switching from magnetized down to up. The white stripe in the middle of the figure, near zero field, indicates a small field region in which data was not taken.

This current-induced switching behavior is characteristic of a spin-orbit torque driven effective field mechanism \cite{Emori}. As $\beta$-Ta has a negative spin-Hall coefficient, charge current flow in the $\textbf{x}$ direction leads to a spin accumulation in the $+\textbf{y}$ direction at the interface with the CoFeB, as indicated schematically in Fig.~\ref{fig_circuit}(d). This leads to a torque on the magnetization also in the $\textbf{y}$ direction which is equivalent to a spin-torque effective field in the direction $\textit{\textbf{H}}_\textrm{ST} \propto \textbf{m} \times \textbf{y}$, where the spin-torque is then proportional to  $\pmb{\mathit{\tau}}_\textrm{ST} \propto \textbf{m} \times \textit{\textbf{H}}_\textrm{ST}$.

It is thus clear that the magnetization must have a component in the $\textbf{x}$ direction for there to be a spin-torque effective field in the $\textbf{z}$ direction to drive magnetization switching (i.e. $m_z \rightarrow -m_z$). An applied field in the $\textbf{x}$ direction leads to a canting of the magnetization in the $\textbf{x}$ direction and thus a preferred switching sense for a given current polarity. Thus the scalar product $\textbf{\textit{J}}\cdot \textbf{\textit{H}}$ determines the sense of the switching: $\textbf{\textit{J}}\cdot \textbf{\textit{H}}>0$ leads to magnetization down to up switching and    $\textbf{\textit{J}}\cdot \textbf{\textit{H}}<0$ leads to magnetization up to down switching, as seen in Fig.~\ref{fig_DC}. The sense of switching would be reversed if Ta was replaced with a material having positive sign of spin Hall coefficient such as Pt  \cite{Liu1}. The critical current density for magnetization reversal is of the order of 5 MA/cm$^2$ for $\mu_0H_x=-100$ mT. We note that there is a small field misalignment ($\lesssim$ 1 deg) in our experiments. The bistable region boundary should have up-down and left-right symmetry. The left-right symmetry is preserved but there is a slight up-down asymmetry. This is associated with a small out-of-plane field component due to a misalignment of the applied field. An estimate of the ratio of the Oersted field to current density 0.5 nm above the center of an infinitely long slab with thickness 5 nm and width 4 $\mu$m is $\approx 0.05$ mT/(MA/cm$^2$), which is two orders of magnitude smaller than the estimated spin-torque effective field to current density ratio ($H_\textrm{ST}/J$, estimated below) and therefore neglected in our analysis.

The decrease of the threshold current with increasing in-plane field is consistent with spin-torque effective field induced switching. Increasing the in-plane field increases $m_x$ and thus the effective field for a given current density, which leads to a reduction of the switching current. However, the decrease in the threshold current near zero field is not expected. We suspect that this is associated with the remanent field from the projection electromagnet (GMW Model 5201) used in this experiment because similar measurements in a conventional dipole magnet does not show a reduced threshold current near zero field.

\begin{figure}[!t]
\centering
\includegraphics[width=3.4in]{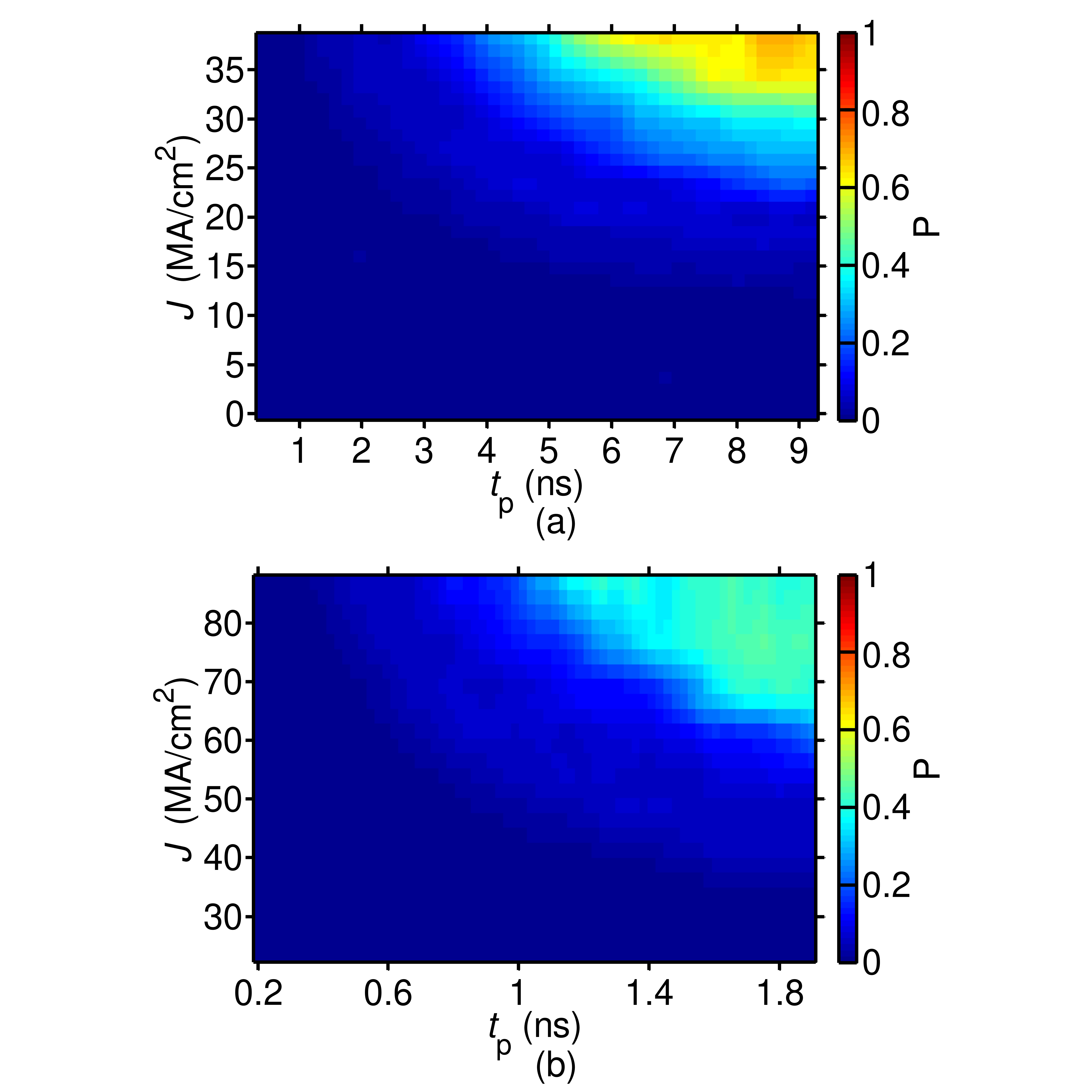}
\caption{Switching probability as a function of pulse duration ($t_{p}$) and amplitude ($J$)  in an applied field of $\mu_0 H_x= -100$ mT. (a) $t_{p}$ from 0.4 to 10 ns and $J$ from 0 to 38.2 MA/cm$^2$. (b) $t_p$ from  0.2 to 1.9 ns and $J$ from 22.1 to 83.5 MA/cm$^2$.}
\label{fig_pulse}
\end{figure}

We now turn to results for pulsed currents. Figure \ref{fig_pulse} shows the switching probability as a function of pulse duration ($t_p$) and amplitude ($J$) in an applied field $\mu_0 H_{x}=-100$ mT. For each pulse amplitude and duration, we apply 100 pulses.  The color of each pixel represents the switching probability, which we define as $P=\frac{1}{N}\sum_{i=1}^{N}\left | \frac{R_{\textrm{after}}-R_{\textrm{before}}}{R_0} \right |_i$, where $N$ is the number of pulses applied. With this definition, the measurement of an intermediate Hall resistance (i.e. $R_{\textrm{after}}-R_{\textrm{before}}<R_0$) adds a fractional contribution to the sum. Thus we also analyze the distribution of Hall resistances after the pulse to determine whether partial or full magnetization switching has occurred.

\begin{figure}[!t]
\centering
\includegraphics[width=3.4in]{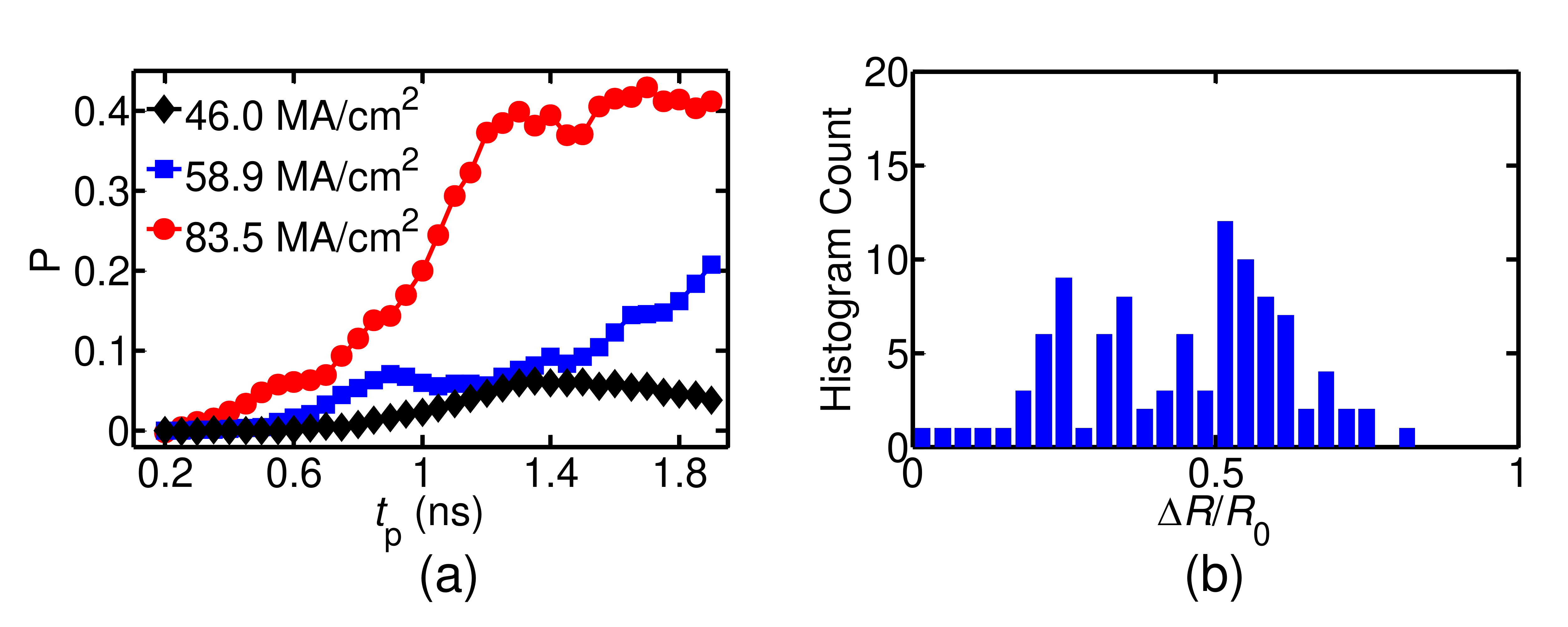}
\caption{(a) Switching probability as a function of pulse duration from 0.2 to 1.9 ns with fixed pulse amplitude. (b) Histogram of the Hall resistance for 100 events with a pulse duration $t_p$ = 1.9 ns and amplitude $J$ = 66.67 MA/cm$^2$.}
\label{fig_cutHistogram}
\end{figure}

Figure \ref{fig_cutHistogram}(a) shows several linescans for fixed pulse amplitudes and varying pulse durations. A switching probability of 0.4 is found with 1 ns duration 83.5 MA/cm$^2$ amplitude pulses. A histogram showing the Hall resistance after a $1.9$ ns duration and 66.67 MA/cm$^2$ amplitude pulse is shown in Fig. \ref{fig_cutHistogram}(b). The observation of intermediate resistance states demonstrates the occurrence of partially switched states, states with up and down magnetized magnetic domains in the current path. 
We observe a switching probability of 0.8 for 10 ns duration $\approx 38$ MA/cm$^2$ amplitude pulses, which is 10 times larger than the switching current density in quasistatic experiments at the same applied field. Additionally, we find partial switched states with 1 ns duration pulses of amplitude 20 times the switching current in quasistatic experiments.

Micromagnetic models and simulations of spin-orbit torque switching in perpendicularly magnetized elements show that it consists of magnetization rotation near the center of the element (where the effective magnetization is the smallest), followed by domain nucleation and propagation \cite{Leecor}. If the injected pulse is of sufficient amplitude and duration for magnetization rotation, domain nucleation and propagation to occur, the result is complete magnetization reversal. In other words, for a given pulse amplitude, the pulse duration for magnetization reversal must be $ t_p \geq t_\mathrm{rotation}+t_\mathrm{nucleation}+t_\mathrm{propagation}$. 
In the quasistatic limit ($t_p  \to \infty$), the magnetization reverses when the current induced spin-orbit torque field is comparable with the domain nucleation field (because the domain propagation field is less than the domain nucleation field in our CoFeB layers). We can estimate the spin-orbit torque effective field at the quasistatic switching current density $J\sim5$ MA/cm$^2$ using $ \textit{\textbf{H}}_\textrm{ST}=\dfrac{\hbar}{2e}\dfrac{J\vert\theta_{\textrm{SH}}\vert}{M_sd}\left(\textbf{m}\times\textbf{y}\right)$, where $\theta_{\textrm{SH}}$ is the spin Hall coefficient and $d$ is the thickness of the magnetic layer.  Taking $\theta_{\textrm{SH}}=-0.15$, we calculate the z-component of $\textit{\textbf{H}}_\textrm{ST}$ to be 0.5 mT, which is comparable with the measured coercive field of 1 mT shown in Fig.~\ref{fig_circuit}(b). In the pulse-current regime, the current density $J$ determines $t_\textrm{rotation}$ and $t_\textrm{propagation}$ because the timescale of magnetization rotation, nucleation and domain propagation depends on the magnitude of spin-torque effective field. Clearly the pulse duration must be sufficiently long for a domain to expand across the layer. If we assume that the reversal is by nucleation and reversed domain expansion we can make a rough estimation of the domain wall velocity from the data in Fig.~\ref{fig_cutHistogram}. From the time for the switching probability to change from 0.1 to 0.2 for 59 MA/cm$^2$ pulses, we find $100$ m/s assuming that a single reversed domain nucleates in the middle of the device and expand isotropically. This estimate is larger than the domain propagation velocities observed in CoFeB \cite{Dafine,Emori}, where velocities of  $\sim 1$ m/s were found at a current density  $\approx 10$ MA/cm$^2$. However, our estimate suggests that domain expansion is rate limiting and the origin of our observation of partial magnetization reversal. 
This may also explain why much higher current densities are found for pulse currents (compared to the quasistatic results); larger current densities are needed to induce faster domain propagation, propagation that can lead a larger region in the current path to reverse during the pulse. 

The current pulse increases the device temperature through Joule heating. Using Fourier's law, assuming a boundary
thermal conductance between the substrate and device of $\kappa = 4$ kW/cm$^2$ and resistivity of $\rho = 200$ $\mu\Omega\cdot$cm, we estimate that the device temperature can increase as much as $\Delta T =J^2 \rho d/\kappa \sim$160~$^{\circ}$C during a $J$ = 80 MA/cm$^2$ current pulse. This increase in temperature is expected to decrease the domain nucleation field and increase the domain propagation velocity. However, the effect of device heating during the pulse on the switching is beyond the scope of our study. We note that the zero temperature threshold for magnetization rotation is $H_\textrm{ST}=M_\textrm{eff}/2$, which gives a zero temperature threshold current density of \mbox{$J_0= 486$ MA/cm$^2$}, which is much larger than what we find in both the pulsed and quasistatic limits, highlighting the possible roles of sample heating during the pulse, thermal fluctuations and defect nucleation sites in spin-orbit torque driven magnetization reversal.

In summary, we have observed full magnetization switching in quasistatic current swept experiments and partially switched states with short current pulses. Our results suggest that the origin of the partial switching is the time required for reversed domain expansion. Pulse switching for 1 ns duration pulses requires current densities 20 times the quasistatic switching threshold. It is clearly of interest to image the magnetization dynamics on short time scales to better understand the magnetization reversal mechanisms. It is also of great interest to optimize materials and element geometries to reduce the switching current densities and time scales for spin orbit torque driven magnetization switching.
\section*{Acknowledgments}
This work was supported by Nanoelectronics Research Initiative (NRI) through the Institute for Nanoelectronics Discovery and Exploration (INDEX). We thank Dr. Tuan Vo and the engineering team at the College of Nanoscale Science and Engineering, Albany, New York for preparing the layer stacks and Dr. Wanjun Jiang and Dr. Suzanne G. E. te Velthuis at Argonne National Labs for conducting the MOKE experiments.

\bibliography{Bib_0707}

\end{document}